\documentclass[varenna]{cimento}
\usepackage{graphicx}  
\usepackage{amssymb}
\usepackage{amsmath}
\usepackage{amsfonts}

\title{Fr\"{o}hlich Polarons From 3D to 0D. \\Concepts and Recent Developments}

\author{J. T. Devreese}

\institute{Theoretische Fysica van de Vaste Stoffen, Universiteit Antwerpen\\ 
CDE, Universiteitsplein, 1, B-2610 Antwerpen, Belgium \\ and eiTT/COBRA interuniversitaire onderzoekschool, 
TU Eindhoven \\ NL-5600 MB Eindhoven, The Netherlands}
\PACSes{PACS 71.38.-k, 63.20.Kr, 71.38.Mx, 71.10.-w, 73.21.-b, 74.20.Mn}

\begin{document}

\maketitle

\begin{abstract}
An overview is presented of the fundamentals of continuum-polaron physics, which provide the basis of the analysis of polaron effects in ionic crystals and polar semiconductors. The present paper deals with \textquotedblleft large\textquotedblright, or \textquotedblleft continuum\textquotedblright, polarons, as described by the Fr\"{o}hlich Hamiltonian. The emphasis is on the polaron optical absorption.
\end{abstract}

\section{The Polaron Concept}

As is generally known, the polaron concept was introduced by Landau in 1933 \cite{Landau}. Gradually 
a distinction was made between polarons in the continuum approximation where long-range electron-lattice 
interaction prevails (\textquotedblleft Fr\"ohlich\textquotedblright polarons) and polarons for which the short-range interaction is essential 
(Holstein, Holstein-Hubbard models). After a period of initial theoretical 
\cite{Landau,LP48,Pekar,Bogolyubov,BT49,Fr54,LLP53,Feynman55} and experimental \cite{B63} works, a first 
international meeting was devoted to the subject in St. Andrews (1965, \cite{KW63}). Ten years later a second 
international meeting -- in which the progress in the field of polarons realized by 1972 was recorded -- 
took place in Antwerpen \cite{Greenbook}. Among the review papers and books covering the subject, we mention 
Refs. \cite{A68,Devreese96,AM96,Calvani}.

Although the present course in Varenna 2005 mainly deals with \textquotedblleft small\textquotedblright polarons (Holstein, Holstein-Hubbard), 
my contribution is devoted to \textquotedblleft large\textquotedblright polarons. The reader is referred to most other papers in the present 
book for an overview of the properties of Holstein related polarons. In the present volume the contributions 
\cite{CalvaniV,CataudellaV,PerroniV,MishchenkoV} also deal with the Fr\"{o}hlich polaron.  

\begin{figure}[tbh]
\begin{center}
\includegraphics[height=.2\textheight]{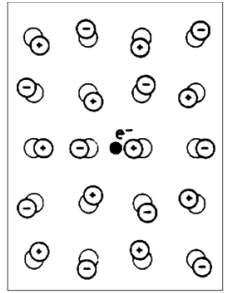}
\end{center}
\caption{Artist view of a polaron. A conduction electron in an ionic crystal
or a polar semiconductor repels the negative ions and attracts the positive
ions. A self-induced potential arises, which acts back on the electron and
modifies its physical properties. (From \cite{Devreese03}.)}%
\label{fig_scheme1}%
\end{figure}

Let us recall the polaron concept for newcomers. A charge placed in a polarizable medium is screened. Dielectric theory describes the phenomenon
by the induction of a polarization around the charge carrier. The induced polarization will follow 
the charge carrier when it is moving through the medium. The carrier together with the induced 
polarization is considered as one entity (see Fig.\thinspace\ref{fig_scheme1}). It was Pekar 
who came up with the term \textit{polaron} \cite{Pekar1946}. The physical properties of a polaron differ 
from those of a band-carrier. A polaron is characterized by its \textit{binding (or self-) energy} 
$E_{0}$, an \textit{effective mass} $m^{\ast}$ and by its characteristic \textit{response} 
to external electric and magnetic fields (e.~g. dc mobility and optical absorption coefficient).

If the spatial extension of a polaron is large compared to the lattice
parameter of the solid, the latter can be treated as a polarizable continuum.
This is the case of a \textit{large, or continuum (Fr\"{o}hlich)} polaron. When the
self-induced polarization caused by an electron or hole becomes of the order
of the lattice parameter, a \textit{small (Holstein)} polaron can arise. 
As distinct from large polarons, small polarons are governed by short-range interactions.

\subsection{The Fr\"{o}hlich Hamiltonian}

Fr\"{o}hlich proposed a model Hamiltonian for the \textquotedblleft 
large\textquotedblright\ polaron through which {its} dynamics is treated
quantum mechanically {(\textquotedblleft  Fr\"{o}hlich
Hamiltonian\textquotedblright)}. The polarization, carried by the longitudinal
optical {(LO)} phonons, is represented by a set of quantum oscillators with
frequency $\omega_{\mathrm{LO}}$, {the long-wavelength LO-phonon frequency,}
and the interaction between the charge and the polarization field is linear in
the field \cite{Fr54}:%
\begin{equation}
H=\frac{\mathbf{p}^{2}}{2m_{b}}+\sum_{\mathbf{k}}\hbar\omega_{\mathrm{LO}%
}a_{\mathbf{k}}^{+}a_{\mathbf{k}}+\sum_{\mathbf{k}}(V_{k}a_{\mathbf{k}%
}e^{i\mathbf{k\cdot r}}+V_{k}^{\ast}a_{\mathbf{k}}^{\dag}e^{-i\mathbf{k\cdot
r}}), \label{eq_1a}%
\end{equation}
where $\mathbf{r}$ is the position coordinate operator of the electron with
band mass $m_{b}$, $\mathbf{p}$ is its canonically conjugate momentum
operator; $a_{\mathbf{k}}^{\dagger}$ and $a_{\mathbf{k}}$ are the creation
(and annihilation) operators for longitudinal optical phonons of wave vector
$\mathbf{k}$ and energy $\hbar\omega_{\mathrm{LO}}$. The $V_{k}$ are the Fourier
components characterising the electron-phonon interaction
\begin{equation}
V_{k}=-i\frac{\hbar\omega_{\mathrm{LO}}}{k}\left(  \frac{4\pi\alpha}%
{V}\right)  ^{\frac{1}{2}}\left(  \frac{\hbar}{2m_{b}\omega_{\mathrm{LO}}%
}\right)  ^{\frac{1}{4}}. \label{eq_1b}%
\end{equation}
The strength of the electron--phonon interaction is {expressed by} a
dimensionless coupling constant $\alpha$, which is defined as:
\begin{equation}
\alpha=\frac{e^{2}}{\hbar}\sqrt{\frac{m_{b}}{2\hbar\omega_{\mathrm{LO}}}%
}\left(  \frac{1}{\varepsilon_{\infty}}-\frac{1}{\varepsilon_{0}}\right).
\label{eq_1c}%
\end{equation}
In this definition, $\varepsilon_{\infty}$ and $\varepsilon_{0}$ are,
respectively, the electronic and the static dielectric constant of the ionic crystal
or polar semiconductor.

In Ref. \cite{Devreese03}) the Fr\"{o}hlich coupling constant is given for a few
solids\footnote{In some cases, due to lack of reliable experimental data to
determine the electron band mass, the values of $\alpha$ are not well
established.}.

{In deriving the form of $V_{k}$, expressions (\ref{eq_1b}) and (\ref{eq_1c}),
it was assumed that (i) the spatial extension of the polaron is large compared
to the lattice parameter of the solid (\textquotedblleft 
continuum\textquotedblright\ approximation), (ii) spin and relativistic
effects can be neglected, (iii) the band-electron has parabolic dispersion,
(iv) in line with the first approximation it is also assumed that the
LO-phonons of interest for the interaction, are the long-wavelength phonons
with constant frequency $\omega_{\mathrm{LO}}$. }

The model, represented by the Hamiltonian (\ref{eq_1a}), (which up to now could
not been solved exactly) has been the subject of extensive investigations,
see, e.~g., Refs.
\cite{Pekar,KW63,A68,Greenbook,Mitra,Devreese96,AM96,Mishchenko2000}. In what
follows the key approaches of the Fr\"{o}hlich-polaron theory are briefly
summarized with indication also of their relevance for the polaron problems in nanostructures.

\section{The Structure of the Polaron Excitation Spectrum and Optical Absorption by Polarons}

\subsection{Optical absorption at weak coupling}
\label{OAweak}

\begin{figure}[tbh]
\begin{center}
\includegraphics[height=.2\textheight]{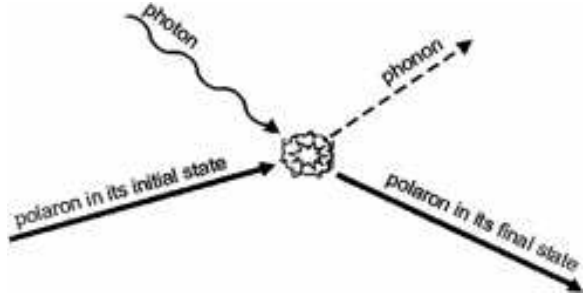}
\end{center}
\caption{Elementary polaron scattering process describing the absorption of an
incoming photon and the generation of an outgoing phonon.}%
\label{fig_scheme}%
\end{figure}

At zero temperature and in the weak-coupling limit, the optical absorption is
due to the elementary polaron scattering process, schematically shown in
Fig.\thinspace\ref{fig_scheme}.

In the weak-coupling limit ($\alpha\ll1$) the polaron absorption coefficient was 
first studied by V. Gurevich, I. Lang and Yu. Firsov \cite{GLF62}. 
Their optical-absorption coefficient is equivalent to a particular approximation 
of the result of J. Tempere and J. T. Devreese (Ref. \cite{TDPRB01}), 
with the dynamic structure factor $S(\mathbf{k},\Omega)$ corresponding 
to the Hartree-Fock approximation (see also \cite{Mahan}, p. 585).

The result of Ref. \cite{TDPRB01} gives the Fr\"ohlich polaron optical absorption coefficient 
for a many-polaron gas (in terms of the dynamic structure factor $S(\mathbf{k},\Omega)$) exactly 
to order $\alpha$. The real part of the optical conductivity of the $N$-polaron system for any $N$ 
is obtained in Ref. \cite{TDPRB01} in an intuitively transparent form
\begin{equation}
\mathrm{Re}[\sigma(\Omega)]=\frac{n}{\hbar\Omega^{3}}\frac{e^{2}}{m_{b}^{2}}%
\sum_{\mathbf{k}}k_{x}^{2}|V_{\mathbf{k}}|^{2}S(\mathbf{k}, \Omega -\omega_{\mathrm{LO}}),  
\label{opticabs}
\end{equation}
where $n=N/V$ is the density of charge carriers and $k_{x}$ is the $x$-component of
the wave vector (the electric field is along the $x$-axis). This approach to the 
many-polaron optical absorption allows one to include the many-body effects in terms 
of the dynamical structure factor 
\begin{equation}
S(\mathbf{k},\Omega )=\int\limits_{-\infty }^{+\infty }\left\langle 
\varphi _{\text{el}}\left| \dfrac{1}{2}\sum\limits_{j,\ell }e^{i\mathbf{k}
\cdot(\mathbf{r}_{j}(t)-\mathbf{r}_{\ell }(0))}\right| \varphi _{\text{el}}
\right\rangle e^{i\Omega t}dt
\label{structurefactor}
\end{equation}
of the {\it electron (or hole) system}. Substituting Eq. (\ref{eq_1b}) in Eq. (\ref{opticabs}), 
the following analytical result was obtained in Ref. \cite{TDPRB01}\footnote{A denominator of the prefactor 
in the Eq. (23) of Ref. \cite{TDPRB01} contain a misprint: there should be $\sqrt{2}$ instead of $2$.}: 
\begin{equation}
\mathrm{Re}[\sigma(\Omega)]=ne^{2}\frac{2\alpha}{3} \frac{1}{\sqrt{2}\pi\Omega^3}
\int\limits_{0}^{+\infty}dk k^2 S(k,\Omega -\omega_{\mathrm{LO}}).  
\label{opticabs1}
\end{equation}

For $N=1$, substituting a $\delta$-function for the dynamical structure factor
\[
S(\mathbf{k},\Omega )\to\delta\left[\frac{\hbar k^{2}}{2m_{b}}-(\Omega -\omega_{\mathrm{LO}})
\right]
\]
in the general expression (\ref{opticabs}), the one-polaron limit Ref. \cite{Devreese72} is retrieved. 
This one-polaron result is derived  \cite{Devreese72} by considering a process, in which the initial state consists of a photon of energy 
$\hbar\Omega$ and a polaron in its ground state, and the final state consists of an emitted LO phonon with energy $\hbar  \omega_{\mathrm{LO}}$
and the polaron, scattered into a state with momentum $\mathbf{k}$ and kinetic energy 
$(\hbar k)^{2}/(2m_{b})=\hbar(\Omega -\omega_{\mathrm{LO}})$. 
The many-polaron result (\ref{opticabs}) is a generalization of the one-polaron picture. The contribution which corresponds to the scattering of a
polaron into the momentum state $\mathbf{k}$ and energy $\hbar(\Omega -\omega_{\mathrm{LO}})$ is now weighed by the dynamical structure factor 
$S(\mathbf{k}, \Omega -\omega_{\mathrm{LO}})$ of the electron (or hole) gas.

\begin{figure}[tbh]
\begin{center}
\includegraphics[height=.25\textheight]{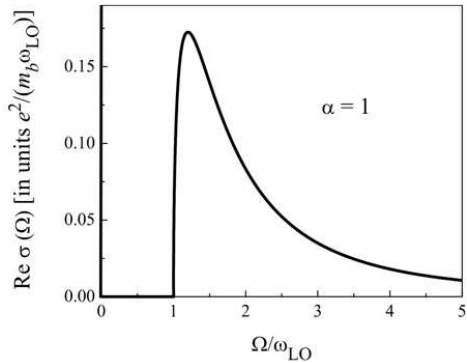}
\end{center}
\caption{Polaron optical conductivity for $\alpha = 1$ in the weak-coupling
approximation. (After \cite{Devreese72}, p. 92.) A $\delta$-like central peak (at $\Omega=0$) is schematically shown by a vertical line.}%
\label{weakabscoef}%
\end{figure}

In the zero-temperature limit, within the weak coupling 
approximation, the real part of the polaron optical conductivity has the following analytic expression 
(see \cite{Devreese72})              :
\begin{equation}
\operatorname{Re}\sigma\left(  \Omega\right)  =\frac{\pi e^{2}}{2m^{\ast}%
}\delta\left(  \Omega\right)  +\frac{2e^{2}}{3m_{b}}\frac{\omega_{\mathrm{LO}}^{3/2}\alpha}
{\Omega^{3}}\sqrt{\Omega-\omega_{\mathrm{LO}}}\Theta\left(
\Omega-\omega_{\mathrm{LO}}\right)  ,\label{Resigma}
\end{equation}
where
\[
\Theta(\Omega-\omega_{\mathrm{LO}})=\left\{
\begin{array}
[c]{lll}%
1 & \mbox{if} & \Omega > \omega_{\mathrm{LO}},\\
0 & \mbox{if} & \Omega < \omega_{\mathrm{LO}}.
\end{array}
\right.
\]
The spectrum of the real part of the polaron optical conductivity (\ref{Resigma}) 
is represented in Fig. \ref{weakabscoef}. The absorption coefficient for absorption 
of light with frequency $\Omega$ by a gas of free polarons 
for $\alpha\longrightarrow0$ takes the form
\begin{equation}
\Gamma_{p}(\Omega)=)=\frac{n_0}{\varepsilon_{0}c n}\frac{\pi e^{2}}{2m^{\ast}%
}\delta\left(  \Omega\right) 
+\frac{1}{\varepsilon_{0}c n}\frac{2ne^{2}\alpha
\omega_{\mathrm{LO}}^{2}}{3m_{b}\Omega^{3}}\sqrt{\frac{\Omega}{\omega
_{\mathrm{LO}}}-1}\quad\Theta\left(\Omega-\omega_{\mathrm{LO}}\right)  , \label{DHL24}%
\end{equation}
where $\epsilon_{0}$ is the dielectric permittivity of the vacuum, $n$ is the refractive index of the medium, 
$n_0$ is the concentration of polarons. A simple derivation in Ref.\thinspace\cite{DHL1971} using a canonical
transformation method gives the absorption coefficient of free polarons, which
coincides with the second term in the result (\ref{DHL24}). A step function in (\ref{DHL24}) reflects the fact that at zero temperature 
the absorption of light accompanied by the emission of a phonon can occur (i) at $\Omega=0$ and (ii) if the energy of 
the incident photon is larger than that of a phonon ($\Omega>\omega_{\mathrm{LO}}$). 
The second term in the absorption coefficient (\ref{DHL24}) was obtained in Ref. \cite{GLF62} in the low-concentration region.
In the weak-coupling limit, 
according to Eq.\thinspace(\ref{DHL24}), the absorption spectrum consists of a central peak [$\propto \delta(\Omega)$] and 
a \textquotedblleft  one-phonon sideband\textquotedblright. At nonzero temperature, the absorption of a photon can be 
accompanied not only by emission, but also by absorption of one or more phonons.

\begin{figure}[hbt]
\begin{center}
\includegraphics[height=.3\textheight]{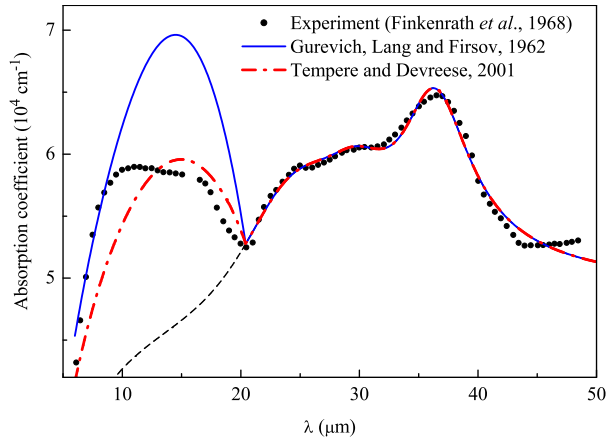}
\end{center}
\caption{ Optical absorption spectrum of a CdO-film with the carrier
concentration $n=5.9\times10^{19}$~cm$^{-3}$ at $T=300$ K. The experimental
data (solid dots) of Ref.~\cite{Finkenrath} are compared to different
theoretical results: with (solid curve) and without (dashed line) the
one-polaron contribution of Ref. \cite{GLF62} and for many polarons
(dash-dotted curve) of Ref. \cite{TDPRB01}. The material parameters of CdO used for calculations: $\alpha =0.74$ \cite{Finkenrath}, $\omega_{{\rm LO}}=490$ cm$^{-1}$ (from the experimental optical absorption
spectrum, Fig. 2 of Ref. \cite{Finkenrath}), $m_b=0.11m_e $ \cite{Landolt}, $\varepsilon \left( 0\right) =21.9,$ $\varepsilon \left( \infty \right) =5.3$ \cite{Landolt}. }%
\label{fig_absCdO}%
\end{figure}

Experimentally, this one-phonon-sideband structure has been observed for free polarons e.g. in the infrared absorption 
spectra of CdO-films, see Fig.\thinspace \ref{fig_absCdO}. In CdO, which is a weakly polar material with $\alpha
\approx0.74$, the polaron absorption band is observed in the spectral region
between 6 and 20 $\mu$m~(above the LO phonon frequency). The difference
between theo\-ry and experiment in the wavelength region where polaron
absorption dominates the spectrum is due to many-polaron effects. 

\subsection{Optical Absorption at Strong Coupling}

The problem of the structure of the large  polaron excitation spectrum 
constituted a central question in the early stages of the development of polaron theory.
The exactly solvable polaron model of Ref. \cite{DThesis} (see Appendix \ref{section-A1}) was used to 
demonstrate the existence of the so-called \textquotedblleft relaxed excited states\textquotedblright of large polarons \cite{DE64}. 
In Ref. \cite{R65}, and after earlier intuitive analysis,  
this  problem was studied using the classical equations of motion and Poisson-brackets. The insight 
gained as a result of those investigations concerning  the structure of the excited polaron states, was 
subsequently used to develop a theory of the optical absorption spectra of polarons. 
The first work was limited to the strong coupling limit \cite{KED69}. Ref. \cite{KED69} is the first work that reveals 
the impact of the internal degrees of freedom of polarons on their optical properties.

The optical absorption of light by free Fr\"ohlich polarons was treated in Ref.~\cite{KED69} using the polaron 
states obtained within the adiabatic strong-coupling approximation. It was argued in Ref.~\cite{KED69}, 
that for sufficiently large $\alpha$ ($\alpha>3$), the (first) relaxed excited state (RES) of a polaron 
is a relatively stable state, which can participate in optical absorption transitions. This idea was necessary 
to understand the polaron optical absorption spectrum in the strong-coupling regime. The following scenario of 
a transition, which leads to a \textit{\textquotedblleft  zero-phonon\textquotedblright\ peak} in the absorption 
by a strong-coupling polaron, can then be suggested. If the frequency of the incoming photon is equal to
$\Omega_{\mathrm{RES}}=0.065\alpha^{2}\omega_{\mathrm{LO}},$
the electron jumps from the ground state (which, at large coupling, is well-characterized by "$s$"-symmetry 
for the electron) to an excited state ("$2p$"), while the lattice polarization in the final state is adapted 
to the "$2p$" electronic state of the polaron. In Ref. \cite{KED69} considering the decay of the RES with 
emission of one real phonon it is argued, that the \textquotedblleft  zero-phonon\textquotedblright\ 
peak can be described using the Wigner-Weisskopf formula valid when the linewidth of that peak is much smaller 
than $\hbar\omega_{\mathrm{LO}}.$

For photon energies larger than
$\Omega_{\mathrm{RES}}+\omega_{\mathrm{LO}},$
a transition of the polaron towards the first scattering state, belonging to the RES, becomes possible. 
The final state of the optical absorption process then consists of a polaron in its lowest RES plus a free phonon. 
A \textquotedblleft  one-phonon sideband\textquotedblright\ then appears in the polaron absorption spectrum. 
This process is called \textit{one-phonon sideband absorption}. The one-, two-, ... $K$-, ... phonon 
sidebands of the zero-phonon peak give rise to a broad structure in the absorption spectrum. It turns out that 
the \textit{first moment} of the phonon sidebands corresponds to the Franck-Condon (FC) frequency
$\Omega_{\mathrm{FC}}=0.141\alpha^{2}\omega_{\mathrm{LO}}.$
To summarize, the polaron optical absorption spectrum at strong coupling is characterized by the following features 
(at $T=0$):
\begin{enumerate}
\item[a)] An intense absorption peak (\textquotedblleft zero-phonon line\textquotedblright) appears, which
corresponds to a transition from the ground state to the first RES at
$\Omega_{\mathrm{RES}}$.
\item[b)] For $\Omega>\Omega_{\mathrm{RES}}+\omega_{\mathrm{LO}}$, a phonon
sideband structure arises. This sideband structure peaks around $\Omega
_{\mathrm{FC}}$.
\end{enumerate}
The qualitative behaviour predicted in Ref.\thinspace\cite{KED69}, namely, an
intense zero-phonon (RES) line with a broader sideband at the high-frequency
side, was confirmed after an all-coupling expression for the polaron optical
absorption coefficient had been studied \cite{DSG72}.

\subsection{Optical Absorption of Continuum Polarons at Arbitrary Coupling}

The polaron absorption coefficient $\Gamma(\Omega)$ of light with frequency $\Omega$ 
at arbitrary coupling was first derived by the present author and collaborators \cite{DSG72} 
(see also \cite{PD1983}). It was represented in the form 
\begin{equation}  
\Gamma(\Omega)=-\frac{1}{n\epsilon_{0}c}\frac{e^{2}}{m_{b}}\frac
{\operatorname{Im}\Sigma(\Omega)}{\left[  \Omega-\operatorname{Re}%
\Sigma(\Omega)\right]  ^{2}+\left[  \operatorname{Im}\Sigma(\Omega)\right]
^{2}}\ . \label{eq:P24-2}%
\end{equation}
This general expression was the starting point for a derivation of the
theoretical optical absorption spectrum of a single Fr\"ohlich polaron at
\textit{all electron-phonon coupling strengths} in
Ref.\thinspace\cite{DSG72}. $\Sigma(\Omega)$ is the so-called \textquotedblleft memory function\textquotedblright, 
which contains the dynamics of the polaron and depends on 
$\Omega$, $\alpha$ and temperature. 
The key contribution of the work in \cite{DSG72} was to introduce $\Gamma(\Omega)$ in the form 
(\ref{eq:P24-2}) and to calculate $\operatorname{Re}\Sigma(\Omega)$, which is essentially a 
(technically not trivial) Kramers--Kronig transform of the more simple function 
$\operatorname{Im}\Sigma(\Omega)$. The function $\operatorname{Im}\Sigma(\Omega)$ had been formally 
derived for the Feynman polaron \cite{FHIP} who studied the polaron mobility $\mu$ from the impedance function, 
i.~e. the static limit 
\[
\mu^{-1}=\lim\limits_{\Omega \to 0} \left(\frac{\operatorname{Im}\Sigma(\Omega)}{\Omega}\right).
\]

The nature of the polaron 
excitations was clearly revealed through this polaron optical absorption obtained in \cite{DSG72,PD1983}. 
It was demonstrated in \cite{DSG72} that the Franck-Condon states for Fr\"ohlich polarons 
are nothing else but a superposition of phonon sidebands. It was also established 
in \cite{DSG72} that a relatively large value of the electron-phonon coupling strength
($\alpha > 5.9$) is needed to stabilise the relaxed excited state of the polaron. It was, further, 
revealed  that at weaker coupling only \textquotedblleft scattering states\textquotedblright of the polaron play a 
significant role in the optical absorption \cite{DSG72,DDG71}.  

\begin{figure}[hptb]
\begin{center}
\includegraphics[width=0.6\textwidth]{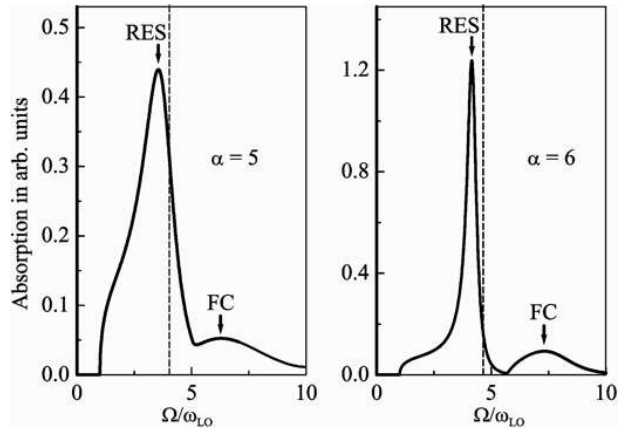}
\end{center}
\caption{Optical absorption spectrum of a Fr\"ohlich polaron for $\alpha=5$ and $\alpha = 6$. 
The RES peak is very intense
compared with the FC peak. The frequency $\Omega/\omega_{\mathrm{LO}}=v$ is
indicated by the dashed lines. The $\delta$-like central peaks (at $\Omega=0$) are schematically 
shown by vertical lines. (From Devreese \textit{et al.} Ref. \cite{DSG72}.)}%
\label{fig_3}%
\end{figure}

In the weak coupling limit, the {optical absorption} spectrum
(\ref{eq:P24-2}) of the polaron is determined by the absorption of radiation
energy, which is reemitted in the form of LO phonons. For $\alpha\gtrsim5.9$,
the polaron can {undergo} transitions toward a relatively stable RES (see
Fig.~\ref{fig_3}). The RES peak in the optical absorption spectrum also has a
phonon sideband-structure, whose average transition frequency can be related
to a FC-type transition. Furthermore, at zero temperature, the optical
absorption spectrum of one polaron exhibits also a zero-frequency
\textquotedblleft  central peak\textquotedblright\ [$\propto\delta(\Omega)$]. For
non-zero temperature, this \textquotedblleft  central peak\textquotedblright%
\ smears out and gives rise to an \textquotedblleft 
anomalous\textquotedblright\ Drude-type low-frequency component of the optical
absorption spectrum. 

For example, in Fig. \ref{fig_3} (from Ref. \cite{DSG72}), the main
peak of the polaron optical absorption for $\alpha=$ 5 at $\Omega
=3.51\omega_{\mathrm{LO}}$ is interpreted as due to transitions to a RES. A
\textquotedblleft  shoulder\textquotedblright\ at the low-frequency side of the
main peak is attributed to one-phonon transitions to polaron-\textquotedblleft 
scattering states\textquotedblright. The broad structure centered at about
$\Omega=6.3\omega_{\mathrm{LO}}$ is interpreted as a FC band. As seen from
Fig. \ref{fig_3}, when increasing the electron-phonon coupling constant to
$\alpha$=6, the RES peak at $\Omega=4.3\omega_{\mathrm{LO}}$ stabilizes. It is
in Ref. \cite{DSG72} that the all-coupling optical absorption spectrum of a
Fr\"{o}hlich polaron, together with the role of RES-states, FC-states and
scattering states, was first presented.
Based on Ref. \cite{DSG72}, it was argued that it is rather Holstein polarons that 
determine the optical properties of the charge carriers in oxides like 
SrTiO$_3$, BaTiO$_3$ \cite{HD75}, while large weak coupling polarons could 
be identified e.~g. in CdO \cite{DHL71}. 
The Fr\"ohlich coupling constants of polar semiconductors and ionic crystals are
generally too small to allow for a static \textquotedblleft RES\textquotedblright. 
In Ref. \cite{ELG95} the RES-peaks of 
Ref. \cite{DSG72} were involved to explain the optical absorption spectrum 
of Pr$_2$NiO$_{4.22}$. Further study of the spectra of  \cite{ELG95} is called for.
The RES-like resonances in $\Gamma(\Omega)$, Eq. (\ref{eq:P24-2}), due to the zero's of $\Omega-\operatorname{Re}%
\Sigma(\Omega)$, can effectively be displaced to smaller polaron coupling by applying an external magnetic field $B$, 
in which case the contribution for what is formally a \textquotedblleft RES-type resonance'' becomes $\Omega-\omega_c-\operatorname{Re}%
\Sigma(\Omega)=0$ ($\omega_c=eB/m_bc$ is the cyclotron frequency). Resonances in the magnetoabsorption governed 
by this contribution have been widely observed and analysed 
\cite{PD86} to \cite{miu97} (see also Subsection \ref{magnetopolarons} below).

\begin{figure}[hptb]
\begin{center}
\includegraphics[width=1.0\textwidth]{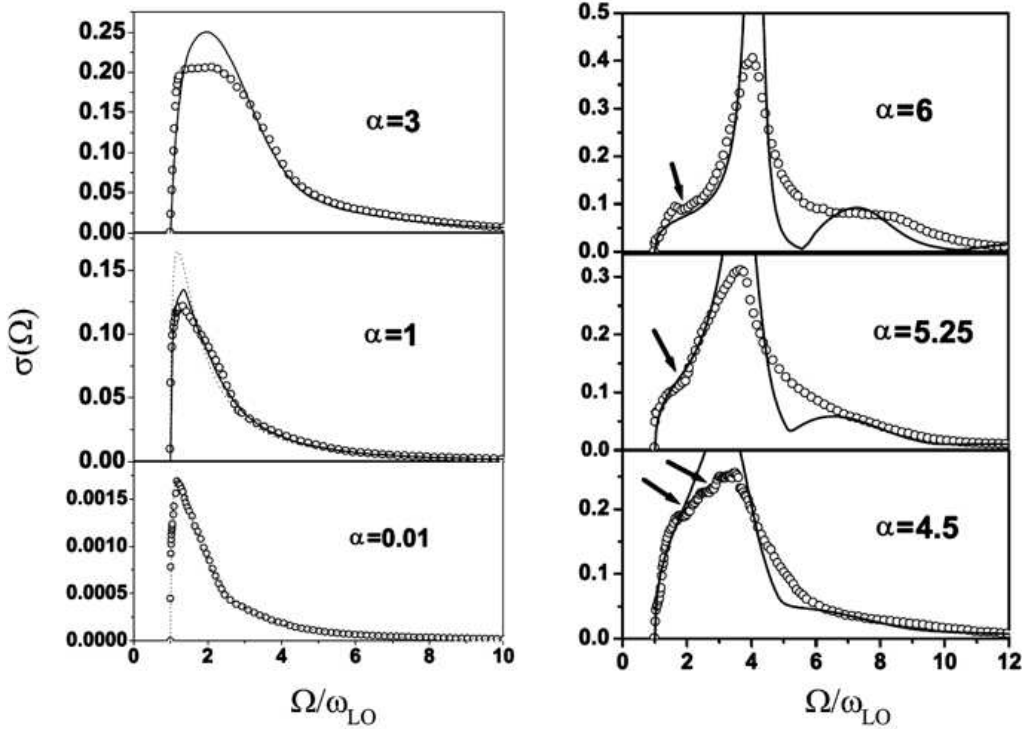}
\end{center}
\caption{{\emph{Left-hand panel}}: Monte Carlo optical conductivity spectra of
one Fr\"{o}hlich polaron for the weak-coupling regime (open circles) compared to the
second-order perturbation theory (dotted lines) for $\alpha=0.01$ and
$\alpha=1$ and to the analytical DSG calculations \cite{DSG72} (solid lines).
{\emph{Right-hand panel}}: Monte Carlo optical conductivity spectra for the
intermediate coupling regime (open circles) compared to the analytical DSG
approach \cite{DSG72} (solid lines). Arrows point to the two- and three-phonon
thresholds. (From Ref.\thinspace\cite{Mishchenko2003}.)}%
\label{fig_4}%
\end{figure}

Recent interesting numerical calculations of the optical
conductivity for the Fr\"{o}hlich polaron performed within the Diagrammatic
Quantum Monte Carlo (DQMC) method \cite{Mishchenko2003}, see Fig. \ref{fig_4}, fully
confirm the essential analytical results derived in Ref. \cite{DSG72} 
for $\alpha\lesssim 3.$ In the intermediate coupling
regime $3<\alpha<6,$ the low-energy behavior and the position of the RES-peak
in the optical conductivity spectrum of Ref. \cite{Mishchenko2003} follow
closely the prediction of Ref. \cite{DSG72}. There are some minor qualitative
differences between the two approaches in the intermediate coupling regime: in
Ref. \cite{Mishchenko2003}, the dominant (\textquotedblleft 
RES\textquotedblright) peak is less intense in the Monte-Carlo numerical
simulations and the second (\textquotedblleft FC\textquotedblright) peak
develops less prominently. The following qualitative differences exist
between the two approaches in the strong coupling regime: in
Ref. \cite{Mishchenko2003}, the dominant peak broadens and the second peak does
not develop, giving instead rise to a flat shoulder in the optical
conductivity spectrum at $\alpha=6.$  These  qualitative
differences from the optical absorption spectrum of Ref. \cite{DSG72} can be
tentatively attributed to the optical processes with participation of two
\cite{Goovaerts73} or more phonons. The above differences can arise also due
to the fact that, starting from the Feynman polaron model, one-phonon processes are
assigned more oscillator strength and the RES of \cite{DSG72} therefore tends to be more stable as
compared to the Monte-Carlo result. The comparison of the analytical and numerical results for the 
optical absorption of the Fr\"{o}hlich polarons shown in Fig. \ref{fig_4}
deserves further study. An independent numerical simulation is called for.

\subsection{Sum Rules for the Optical Conductivity Spectra}

In this section the sum rules for the optical conductivity spectra of Fr\"{o}hlich polarons 
obtained within the DSG approach \cite{DSG72} are compared with those found from the
DQMC results \cite{Mishchenko2003}. The values of the
polaron effective mass for the Monte Carlo approach are taken from Ref.
\cite{Mishchenko2000}. In Tables \ref{tab3} and \ref{tab4}, I show the polaron ground-state $E_{0}$ 
and the following parameters calculated using the optical conductivity spectra:%
\begin{align}
M_{0}  &  \equiv\int_{1}^{\omega_{\max}}\operatorname{Re}\sigma\left(
\omega\right)  d\omega,\label{1}\\
M_{1}  &  \equiv\int_{1}^{\omega_{\max}}\omega\operatorname{Re}\sigma\left(
\omega\right)  d\omega, \label{2}%
\end{align}
where $\omega_{\max}$ is the upper value of the frequency available from Ref.
\cite{Mishchenko2003},%
\begin{equation}
\tilde{M}_{0}\equiv\frac{\pi}{2m^{\ast}}+\int_{1}^{\omega_{\max}%
}\operatorname{Re}\sigma\left(  \omega\right)  d\omega, \label{zm}%
\end{equation}
where $m^{\ast}$ is the polaron mass, the optical conductivity is calculated
in units $n_{0}e^{2}/(m_{b}\omega_{\mathrm{LO}}),$ $m^{\ast}$ is
measured in units of the band mass $m_{b}$, and the frequency is measured in
units of $\omega_{\mathrm{LO}}$. The values of $\omega_{\max}$ are:
$\omega_{\max}=10$ for $\alpha=0.01,$ 1 and 3, $\omega_{\max}=12$ for
$\alpha=4.5,$ 5.25 and 6, $\omega_{\max}=18$ for $\alpha=6.5,$ 7 and
8.

\begin{table}
\caption{Polaron parameters $M_0, M_1, {\tilde M}_0$ obtained from the diagrammatic Monte
Carlo results}
\label{tab3}
\begin{tabular}
[c]{llllll}\hline
$\alpha$ & $M_{0}^{\left(  \mathrm{MC}\right)  }$ & $m^{\ast\left(
\mathrm{MC}\right)  }$ & $\tilde{M}_{0}^{\left(  \mathrm{MC}\right)  }%
$&$M_{1}^{\left(  \mathrm{MC}\right)  }/\alpha$ & $E_{0}^{\left(  \mathrm{MC}%
\right)  }$\\\hline
$0.01$ & $0.00249$ & $1.0017$ & $1.5706$&$0.634$ & $-0.010$\\
$1$ & $0.24179$ & $1.1865$ & $1.5657$&$0.65789$ & $-1.013$\\
$3$ & $0.67743$ & $1.8467$ & $1.5280$&$0.73123$ & $-3.18$\\
$4.5$ & $0.97540$ & $2.8742$ & $1.5219$&$0.862$ & $-4.97$\\
$5.25$ & $1.0904$ & $3.8148$ & $1.5022$&$0.90181$ & $-5.68$\\
$6$ & $1.1994$ & $5.3708$ & $1.4919$&$0.98248$ & $-6.79$\\
$6.5$ & $1.30$ & $6.4989$ & $1.5417$&$1.1356$ & $-7.44$\\
$7$ & $1.3558$ & $9.7158$ & $1.5175$&$1.2163$ & $-8.31$\\
$8$ & $1.4195$ & $19.991$ & $1.4981$&$1.3774$ & $-9.85$\\\hline
\end{tabular}
\end{table}

\begin{table}
\caption{Polaron parameters $M_0, M_1, {\tilde M}_0$ obtained within the path-integral
approach}
\label{tab4}
\begin{tabular}
[c]{llllll}\hline
$\alpha$ & $M_{0}^{\left(  \mathrm{DSG}\right)  }$ & $m^{\ast\left(
\mathrm{Feynman}\right)  }$ & $\tilde{M}_{0}^{\left(  \mathrm{DSG}\right)  }%
$&$M_{1}^{\left(  \mathrm{DSG}\right)  }/\alpha$ & $E_{0}^{\left(
\mathrm{Feynman}\right)  }$\\\hline
$0.01$ & $0.00248$ & $1.0017$ & $1.5706$&$0.633$ & $-0.010$\\
$1$ & $0.24318$ & $1.1957$ & $1.5569$&$0.65468$ & $-1.0130$\\
$3$ & $0.69696$ & $1.8912$ & $1.5275$&$0.71572$ & $-3.1333$\\
$4.5$ & $1.0162$ & $3.1202$ & $1.5196$&$0.83184$ & $-4.8394$\\
$5.25$ & $1.1504$ & $4.3969$ & $1.5077$&$0.88595$ & $-5.7482$\\
$6$ & $1.2608$ & $6.8367$ & $1.4906$&$0.95384$ & $-6.7108$\\
$6.5$ & $1.3657$ & $9.7449$ & $1.5269$&$1.1192$ & $-7.3920$\\
$7$ & $1.4278$ & $14.395$ & $1.5369$&$1.2170$ & $-8.1127$\\
$8$ & $1.4741$ & $31.569$ & $1.5239$&$1.4340$ & $-9.6953$\\\hline
\end{tabular}%
\end{table}

The parameters corresponding to the Monte Carlo calculation are obtained using
the numerical data kindly provided to the author by A. Mishchenko \cite{Mishchenko2005}. 
The optical conductivity obtained within  
the path-integral approach of Ref.~\cite{DSG72} exactly satisfies the sum rule \cite{SR}%
\begin{equation}
\frac{\pi}{2m^{\ast}}+\int_{1}^{\infty}\operatorname{Re}\sigma\left(
\omega\right)  d\omega=\frac{\pi}{2}. \label{sr}%
\end{equation}
Since the optical conductivity obtained from the Monte Carlo results \cite{Mishchenko2003} is known only 
within a limited interval of frequencies $1< \omega < \omega_{\rm max}$, we calculate the integral 
in Eq. (\ref{zm}) for the DSG-approach \cite{DSG72} over the same frequency interval as for the Monte Carlo
results \cite{Mishchenko2003}. 

The comparison of the resulting zero frequency moments $\tilde{M}_{0}^{\left(  \mathrm{MC}\right)  }$ and
$\tilde{M}_{0}^{\left(  \mathrm{DSG}\right)  }$ with each other and with the
value $\pi/2=1.5707963...$ corresponding to the right-hand-side of the sum rule (\ref{sr})
shows that the difference $\left\vert \tilde{M}_{0}^{\left(  \mathrm{MC}\right)  }-\tilde
{M}_{0}^{\left(  \mathrm{DSG}\right)  }\right\vert $ on the interval $\alpha \leq 8$ 
is smaller than the absolute value of the contribution 
of the \textquotedblleft  tail\textquotedblright\ of the optical conductivity for
$\omega>\omega_{\max}$ to the integral in the sum rule (\ref{sr}):
\begin{equation}
\int_{\omega_{\rm max}}^{\infty}\operatorname{Re}\sigma ^{\left(  \mathrm{DSG}\right) }\left(
\omega\right)  d\omega \equiv\frac{\pi}{2}-\tilde{M}_{0}^{\left(  \mathrm{DSG}\right)}. \label{sr1}%
\end{equation}
Within the accuracy determined by the neglect of the \textquotedblleft  tail\textquotedblright\ 
of the part of the spectrum for $\omega>\omega_{\max}$, the contribution to the integral in the sum rule 
(\ref{sr}) for the optical conductivity obtained from the Monte Carlo results \cite{Mishchenko2003} {\it agrees 
well} with that for the optical conductivity found within the path-integral approach of Ref.~\cite{DSG72}.
Hence, the conclusion follows that {\it the optical conductivity obtained from the Monte Carlo results \cite{Mishchenko2003} 
satisfies the sum rule (\ref{sr})} within the aforementioned accuracy.

\begin{figure}
[hptb]
\begin{center}
\includegraphics[height=0.4\textheight]{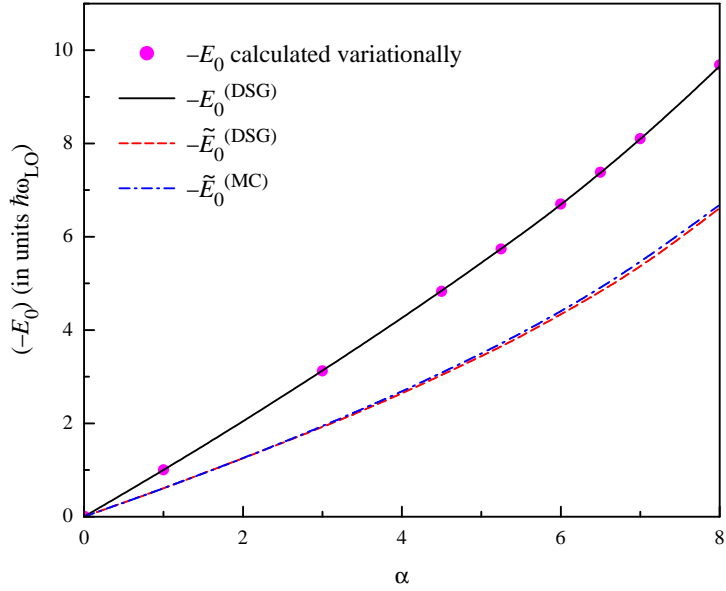}%
\caption{Test of the ground-state theorem for a Fr\"{o}hlich polaron from Ref. \cite{LSD} using different optical 
conductivity spectra, DSG from Ref. \cite{DSG72} and MC from Ref. \cite{Mishchenko2003}. 
The notations are explained in the text.}%
\label{Moments2-f1}%
\end{center}
\end{figure}
We analyze also the fulfillment of the \textquotedblleft LSD\textquotedblright polaron ground-state theorem introduced 
in \cite{LSD}%
\begin{equation}
E_{0}\left(  \alpha\right)  -E_{0}\left(  0\right)  =-\frac{3}{\pi}\int
_{0}^{\alpha}\frac{d\alpha^{\prime}}{\alpha^{\prime}}\int_{0}^{\infty}%
\omega\operatorname{Re}\sigma\left(  \omega,\alpha^{\prime}\right)
d\omega\label{gst}%
\end{equation}
using the first-frequency moments $M_{1}^{\left(  \mathrm{MC}\right)  }$ and
$M_{1}^{\left(  \mathrm{DSG}\right)  }$. The results of this comparison are
presented in Fig. \ref{Moments2-f1}. The dots indicate the polaron
ground-state energy calculated using the Feynman variational principle. The
solid curve is the value of $E_{0}\left(  \alpha\right)  $ calculated
numerically using the optical conductivity spectra and the ground-state
theorem with the DSG optical conductivity \cite{DSG72} for a polaron,%
\begin{equation}
E_{0}^{\left(  \mathrm{DSG}\right)  }\left(  \alpha\right)  \equiv-\frac
{3}{\pi}\int_{0}^{\alpha}\frac{d\alpha^{\prime}}{\alpha^{\prime}}\int
_{0}^{\infty}\omega\operatorname{Re}\sigma^{\left(  \mathrm{DSG}\right)
}\left(  \omega,\alpha^{\prime}\right)  d\omega. \label{e1R}%
\end{equation}
The dashed and the dot-dashed curves are the values obtained using
$M_{1}^{\left(  \mathrm{DSG}\right)  }\left(  \alpha\right)  $ and
$M_{1}^{\left(  \mathrm{MC}\right)  }\left(  \alpha\right)  $, respectively:%
\begin{equation}\tilde{E}_{0}\left(  \alpha\right) 
\equiv-\frac{3}{\pi}\int_{0}^{\alpha}\frac{d\alpha^{\prime}}{\alpha^{\prime}%
}\int_{0}^{\omega_{\max}}\omega\operatorname{Re}\sigma\left(  \omega,\alpha^{\prime}\right)  
d\omega=-\frac{3}{\pi}\int_{0}^{\alpha}d\alpha^{\prime}
\frac{M_{1}\left(  \alpha^{\prime}\right)  }{\alpha^{\prime}}.\label{e2R}
\end{equation}

As seen from the figure, $E_{0}^{\left(  \mathrm{DSG}\right)  }\left(
\alpha\right)  $ to a high degree of accuracy coincides with the variational
polaron ground-state energy. Both $\tilde{E}_{0}^{\left(  \mathrm{DSG}\right)
}\left(  \alpha\right)  $ and $\tilde{E}_{0}^{\left(  \mathrm{MC}\right)
}\left(  \alpha\right)  $ differ from $E_{0}^{\left(  \mathrm{DSG}\right)
}\left(  \alpha\right)  $ due to the integration over the same limited frequency interval.
However, $\tilde{E}_{0}^{\left(  \mathrm{DSG}\right)  }\left(
\alpha\right)  $ and $\tilde{E}_{0}^{\left(  \mathrm{MC}\right)  }\left(
\alpha\right)  $ are very close to each other. Herefrom, the conclusion follows
that for integrals over the finite frequency region characteristic for the
polaron optical absorption (i. e., omitting the \textquotedblleft 
tails\textquotedblright), the function $\tilde{E}_{0}^{\left(  \mathrm{MC}%
\right)  }\left(  \alpha\right)  $ reproduces very well the
function $\tilde{E}_{0}^{\left(  \mathrm{DSG}\right)  }\left(  \alpha\right)
$.%

From the comparison of $\tilde{E}_{0}^{\left(  \mathrm{DSG}\right)
}\left(  \alpha\right)  $ with $\tilde{E}_{0}^{\left(  \mathrm{MC}\right)
}\left(  \alpha\right)  $ it follows that the contribution to the integral in (\ref{e2R})
with the limited frequency region, which approximates the integral in the right-hand side of the 
\textquotedblleft LSD\textquotedblright ground state theorem (\ref{gst}), for the optical conductivity obtained from the Monte Carlo results 
\cite{Mishchenko2003} {\it agrees with a high accuracy} with the corresponding contribution to the integral in 
(\ref{e2R}) for the optical conductivity found within the path-integral approach of Ref.~\cite{DSG72}. 
Because for the path-integral result, the integral $\tilde{E}_{0}^{\left(  \mathrm{DSG}\right)}\left(  \alpha\right)  $  
noticeably differs from  the available integral ${E}_{0}^{\left(  \mathrm{DSG}\right)}\left(  \alpha\right)  $, 
a comparison between the Feynman polaron ground state energy $E_0$ and the integral 
$\tilde{E}_{0}^{\left(  \mathrm{DSG}\right)}\left(  \alpha\right)  $ is not representative. 
Similarly, a comparison between the polaron ground state energy obtained from the Monte Carlo results and 
the integral $\tilde{E}_{0}^{\left(  \mathrm{MC}\right)}\left(  \alpha\right)  $  would require to overcome 
the present limitation for the available optical conductivity spectrum \cite{Mishchenko2003}.    

In conclusion, the Monte Carlo optical conductivity spectrum for higher frequencies than $\omega_{\max}$ of \cite{Mishchenko2003} 
is needed in order to check the fulfillment of the sum rules (\ref{sr}) and (\ref{gst}) with a higher accuracy.

\section{Scaling Relations for Fr\"{o}hlich Polarons}

Several scaling relations connect the polaron self-energy, the effective
mass, the impedance $Z$ and the polaron mobility $\mu $ in 2D to their counterpart in 3D. Those relations 
were introduced in Ref. \cite{PD87} (Peeters and Devreese), they are satisfied at the level of the Feynman polaron 
model and are listed here:
\begin{eqnletter}
\label{scaling_all}
E_{\mathrm{2D}}(\alpha ) &=&\frac{2}{3}E_{\mathrm{3D}}\left( \frac{3\pi }{4}%
\alpha \right),  \label{Escaling} \\
\frac{m_{\mathrm{2D}}^\ast(\alpha)}{m_{\mathrm{2D}}} &=& \frac{m_{\mathrm{3D}%
}^\ast(\frac{3}{4}\alpha)}{m_{\mathrm{3D}}}\ ,  \label{eq:P24-3} \\
Z_{\mathrm{2D}}(\alpha ,\Omega ) &=&Z_{\mathrm{3D}}\left( \frac{3\pi }{4}\alpha
,\Omega \right), \\
\mu _{\mathrm{2D}}(\alpha ) &=& 
\mu _{\mathrm{3D}}\left( \frac{3\pi }{4}\alpha
\right).  \label{MU}
\end{eqnletter}
In Eq. (\ref{eq:P24-3}), {$m_{\mathrm{2D}}^\ast$ ($m_{\mathrm{3D}}^\ast$)
and $m_{\mathrm{2D}}$ ($m_{\mathrm{3D}}$) are, respectively, the polaron and
the electron-band masses in 2D (3D)}. Expressions (\ref{scaling_all}) provide a straightforward link between polaron 
characteristics in 3D with those in 2D. The scaling relations imply that polaron effects may be observable 
in 2 dimensions at substantially lower electron-phonon coupling than in 3 dimensions. Against this background, 
an extensive analysis of polaronic phenomena was undertaken for GaAs heterostructures, quantum wells, superlattices, 
2D oxide-structures, shallow donors (\textquotedblleft bound polarons\textquotedblright) and also for quantum dots, see 
e.~g. Refs. \cite{7} to \cite{KFBD2004} and the reviews \cite{Devreese96,Devreese03}. 

\clearpage 
\begin{figure}[tbh]
\begin{center}
\includegraphics[height=0.8\textheight]{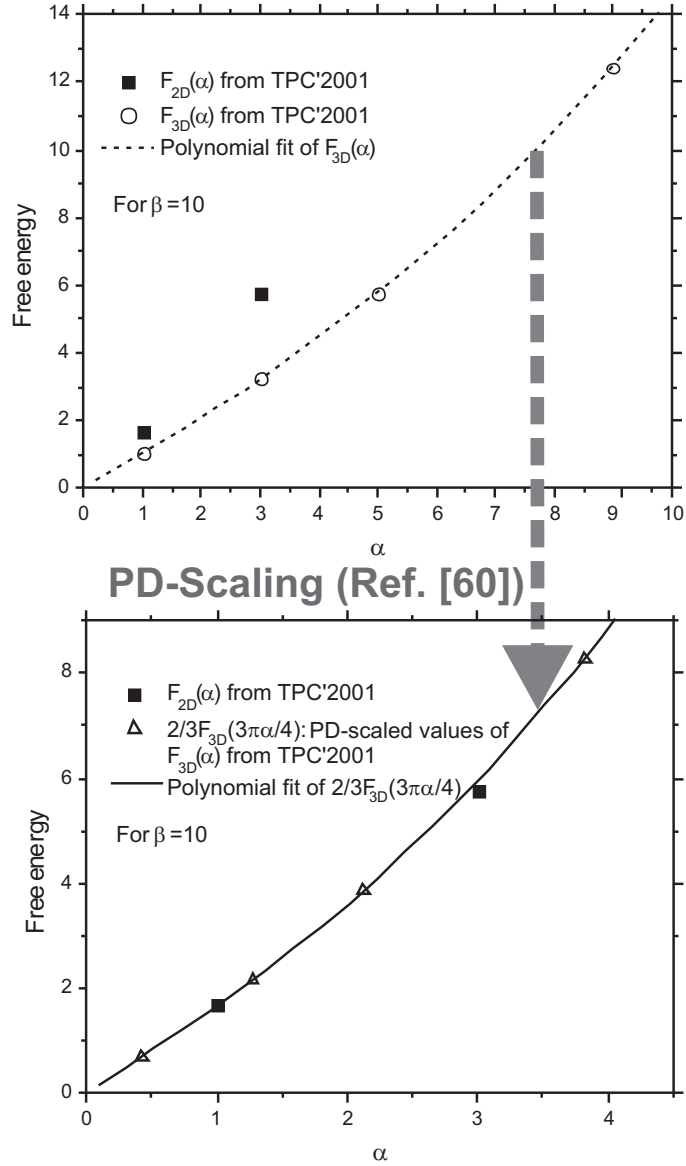}
\end{center}
\caption{\textit{Upper panel: }Polaron free energy in 3D
(squares) and 2D (open circles) obtained by TPC'2001 \cite{TPC} for $\beta
=10$. The data for $F_{3D}\left(  \alpha\right)  $ are interpolated\ using a
polynomial fit to the available four points (dotted line). \textit{Lower
panel:} Demonstration of the PD-scaling cf. Peeters-Devreese \cite{PD87}: the polaron
free energy in 2D obtained by TPC'2001 \cite{TPC} for $\beta=10$ (squares) is
very close to the \textbf{PD-scaled} according to Peeters-Devreese \cite{PD87}
values of the polaron free energy in 3D from TPC'2001 for $\beta=10$ (open
triangles). The data for $\frac{2}{3}F_{3D}\left(  \frac{3\pi\alpha}%
{4}\right)  $ are interpolated\ using a polynomial fit to the available four
points (solid line).}%
\label{ScComp}%
\end{figure}
\clearpage

\subsection{Check of the scaling relation for the path integral Monte Carlo
results for the polaron free energy}

The fulfillment of the PD-scaling relation \cite{PD87} is now checked for the
path integral Monte Carlo results \cite{TPC} for the polaron free energy.
The path integral Monte Carlo results of Ref. \cite{TPC} for the polaron free
energy in 3D and in 2D were given for a few values of temperature and for some
selected values of $\alpha.$ For a check of the scaling relation, the values
of the polaron free energy at $\beta=10$ are taken from Ref. \cite{TPC} in 3D
(Table I of Ref. \cite{TPC}, for four values of $\alpha$) and in 2D (Table II of Ref. \cite{TPC}, 
for two values of $\alpha$) and plotted in Fig. \ref{ScComp}, upper panel, with squares and open
circles, correspondingly.
In Fig. \ref{ScComp}, lower panel, the available data for the free energy from
Ref \cite{TPC} are plotted in the following form \textit{inspired by the
l.h.s. and the r.h.s parts of Eq. (\ref{Escaling})}: $F_{2D}\left(  \alpha\right)  $
(squares) and $\frac{2}{3}F_{3D}\left(  \frac{3\pi\alpha}{4}\right)  $(open
triangles). As follows from the figure, t\textit{he path integral Monte Carlo
results for the polaron free energy in 2D and 3D very closely follow the
PD-scaling relation of the form given by Eq. (\ref{Escaling}):}%
\begin{equation}
F_{2D}\left(  \alpha\right)  \equiv\frac{2}{3}F_{3D}\left(  \frac{3\pi\alpha
}{4}\right)  . \label{F}%
\end{equation}
\subsection{Magnetoabsorption spectra of polarons} 
\label{magnetopolarons}

\begin{figure}[tbh]
\begin{center}
\includegraphics[height=0.4\textheight]{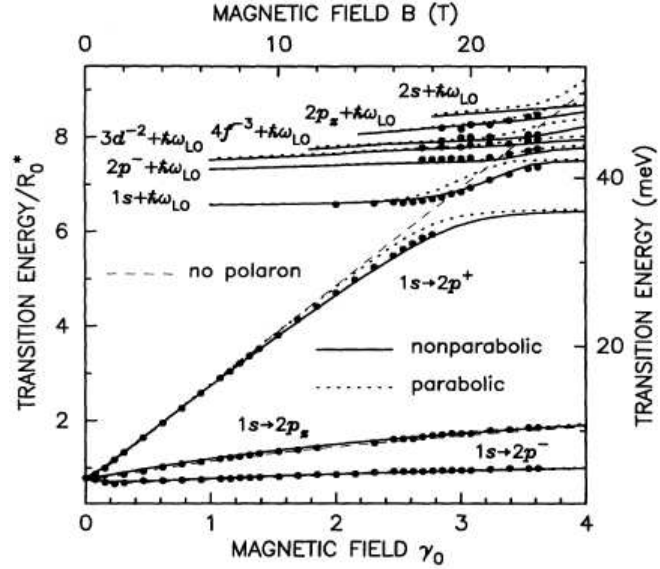}
\end{center}
\caption{The $1s \to 2p^{\pm},2p_z$ transition energies as a function of a magnetic field for a donor in GaAs. 
We compare our theoretical results for the following cases: (a) without the effect of polaron and band 
nonparabolicity (thin dashed curves); (b) with polaron correction (dotted curves); (c) including the effects 
of polaron and band nonparabolicity (solid curves) to the experimental data of Ref. \cite{Cheng} (solid dots).
(From Ref. \cite{SPD93}.)}
\label{magnetoabs}%
\end{figure}

The results on the polaron optical absorption \cite{DSG72,Devreese72} paved the way for an 
all-coupling path-integral based theory of magneto-optical absorption by polarons (see Ref. \cite{PD86}, 
F. Peeters and J. T. Devreese). This work was i.~a. motivated by the insight that magnetic fields can stabilise 
the relaxed excited polaron states, so that information on the nature of relaxed excited states might be gained from the 
cyclotron resonance of polarons. A quantitative interpretation of the 
high-precision cyclotron resonance experiment on AgBr and AgCl in a range of magnetic fields up to 15 T \cite{Hodby} 
by the theory of Ref. \cite{PD86} provided one of the most convincing and clearest demonstrations of polaron 
features in solids. For a review of the subsequent developments in the field of polaron cyclotron resonance, 
the reader is referred to  Refs. \cite{PWD92,SPD93,Grynberg1996,prl97,Peeters2003,Devreese96,Devreese03} and 
references therein.  The energy spectra of such polaronic systems as shallow donors (\textquotedblleft bound polarons\textquotedblright), 
e. g., the D$_0$ and D$^-$ centres, constitute  the most complete and detailed 
polaron spectroscopy realised in the literature, see for example Fig. \ref{magnetoabs}. 

\section{Many-Polaron Systems}

\subsection{Ground State and the Optical Absorption Spectra of Many-Polaron Systems}

The earliest studies on many polarons concerned their role in heavily 
doped polar semiconductors \cite{Mahan1972,Sernelius1987,Sherman2003}.
The possibility that polarons and bipolarons play a role in high-$T_{c}$
superconductors has renewed interest in the physical properties of
many-polaron systems and, in particular, in their optical properties.
Theoretical treatments have been extended from one-polaron to many-polaron
systems both of Fr\"{o}hlich polarons (see, for example, 
\cite{Mahan1966,LDB77,Fratini2000,Devreese03,Bassani,Fratini2002,Perroni2004,Iadonisi2005}) 
and of Holstein polarons (see, e. g., \cite{Berger1995,Fratini2001,Takada2003}).

For the weak-coupling regime, which is realized in most polar
semiconductors, the ground-state energy of a gas of interacting large 
polarons was derived in Ref.\,\cite{LDB77} by introducing a variational wave
function:
\begin{equation}
\left| \psi_{\text{LDB}}\right\rangle =U\left| \phi\right\rangle \left|
\varphi_{\text{el}}\right\rangle ,  \label{psiLDB}
\end{equation}
where $\left| \varphi_{\text{el}}\right\rangle $ represents the ground-state
many-body wave function for the electron (or hole) system, $\left|
\phi\right\rangle $ is the phonon vacuum and $U$ is a many-body unitary
operator that determines a canonical transformation for a fermion gas
interacting with a boson field:
\begin{equation}
U=\exp\left\{ \sum_{j=1}^{N}\sum_{\mathbf{k}}\left( f_{\mathbf{k}}a_{\mathbf{%
k}}e^{i\mathbf{k}\cdot\mathbf{r}_{j}}-f_{\mathbf{k}}^{\ast }a_{\mathbf{k}%
}^{+}e^{-i\mathbf{k}\cdot\mathbf{r}_{j}}\right) \right\},  \label{U}
\end{equation}
where $\mathbf{r}_{j}$ represent the position vectors of the $N$ constituent
electrons (or holes); $a_{\mathbf{k}}^{+},a_{\mathbf{k}}$ denote the
creation and annihilation operators for LO phonons with wave vector $\mathbf{%
k}$. The $f_{\mathbf{k}}$ were determined variationally \cite{LDB77}. It may
be emphasized that the expression (\ref{U})
constitutes -- especially in its implementation -- a
nontrivial extension of the one-particle approximation to a many-body system.
An advantage of the many-polaron canonical transformations introduced in
Ref.\,\cite{LDB77} for the calculation of the ground state energy of a polaron
gas is that, to order $\alpha$, the many-body effects are entirely contained in 
the static structure factor of the electron (or hole) system, 
which appears in the analytical expression for the energy. 
The minimum of the total ground-state energy per particle 
as a function of the electron density for a polaron gas is shown \cite{LDB77} 
to lie at lower value of the density than that for the electron gas.

\begin{figure}[ptbh]
\begin{center}
\includegraphics[width=0.8\textwidth]{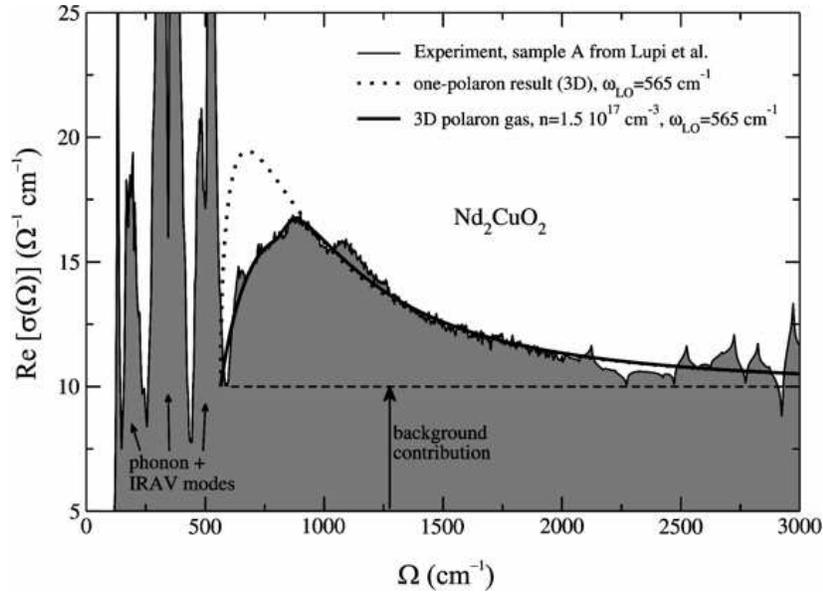}
\end{center}
\caption{The infrared absorption of Nd$_{2}$CuO$_{2-\delta}$ ($\delta<0.004$)
is shown as a function of frequency, up to 3000 cm$^{-1}$. The experimental
results of Calvani and co-workers \cite{calva2} is represented by the thin
black curve and by the shaded area. The so-called `d-band' rises in intensity
around 600 cm$^{-1}$ and increases in intensity up to a maximum around 1000
cm$^{-1}$. The dotted curve shows the single polaron result. The full black
curve represents the theoretical results obtained in the present work for the
interacting many-polaron gas with $n_{0}=1.5\times10^{17}$ cm$^{-3}$,
$\alpha=2.1$ and $m_{b}=0.5$ $m_{e}$. (Recently, a numerical inaccuracy was {identified} 
in the fitting procedure, concerning {the} 
strength of the plasmon branch, so that the parameter values given above may not 
be the optimal parameter values, see Ref.\,\protect\cite{Hameeuw2004}.) 
(From Ref. \cite{TDPRB01}.)}%
\label{polabs1}%
\end{figure}

\begin{figure}[tbp]
\begin{center}
\includegraphics[height=.3\textheight]{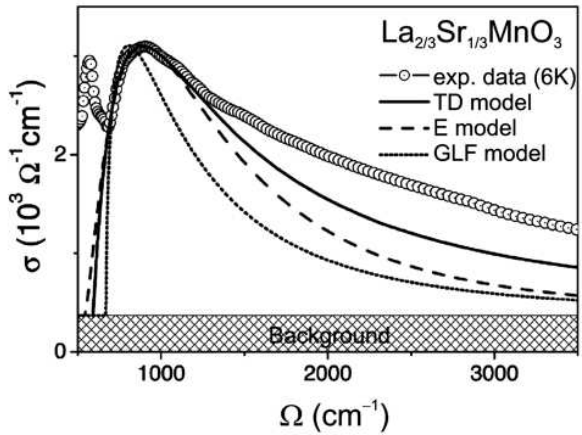}
\end{center}
\caption{Comparison of the measured mid-infrared optical conductivity in 
La$_{2/3}$Sr$_{1/3}$MnO$_{3}$ at $T=6$ K to that given by several model calculations 
for $m=3m_{\mathrm{e}}$, $\alpha$ of the order of 1 and $n_{0}=6\times 10^{21}$ cm$^{-3}$. 
The one-polaron approximations [the weak-coupling approach by V. L. Gurevich, I. G. Lang, and Yu.~A.~Firsov 
{\cite{GLF62}} (GLF model) and the phenomenological approach by D. Emin {\cite{Emin1993}} (E model)] 
lead to narrower polaron peaks than a peak with maximum at \protect\linebreak $\omega \sim 900$ cm$^{-1}$ 
given by the many-polaron treatment by J. Tempere and J.~T.~Devreese (TD model) of Ref.\,\protect\cite{TDPRB01}. 
(After {Ch. Hartinger {\it et al.},} Ref.\,\protect\cite{Hartinger2004}.)}
\label{polabs2}
\end{figure}

In Ref. \cite{TDPRB01}, the optical absorption coefficient of an interacting 
many-polaron gas has been derived starting from the many-polaron canonical
transformations and the variational many-polaron wave function introduced in
Ref.~\cite{LDB77}. The real part of the optical conductivity of the many-polaron system 
to order $\alpha$ is obtained in terms of the dynamical structure factor in Ref. \cite{TDPRB01} 
in the form (\ref{opticabs}), which has been discussed in the Subsection \ref{OAweak}.  
The experimental peaks in the mid-infrared optical absorption
spectra of  (Fig.\thinspace\ref{polabs1}) and manganites (Fig.\thinspace\ref{polabs2})
have been interpreted within this theory, using (\ref{opticabs}).
The experimental optical absorption spectrum (up to 3000 cm$^{-1}$) of Nd$_{2}$CuO$_{2-\delta}$ ($\delta<0.004$), 
obtained by P. Calvani and co-workers \cite{calva2}, is shown in Fig. \ref{polabs1} (shaded area) together with 
the theoretical curve obtained by the method of Ref. \cite{TDPRB01} (full, bold curve) and, for reference, with the 
one-polaron optical absorption result (dotted curve). At lower frequencies (600-1000 cm$^{-1}$) a marked difference 
between the single polaron optical absorption and the many-polaron result is manifest. The experimental 
\emph{d}-band can be clearly identified, rising in intensity at about 600 cm$^{-1}$, peaking around 1000 cm$^{-1}$, 
and then decreasing in intensity above that frequency. At a density of $n_{0}=1.5\times10^{17}$ cm$^{-3}$, we found 
compelling agreement between our theoretical predictions and the experimental curve. The authors of 
Ref.\,\protect\cite{Hartinger2004} interpret their data on La$_{2/3}$Sr$_{1/3}$Mn)$_3$ with Eq.~(\ref{opticabs}) 
\cite{TDPRB01}, adapted for an on-site Hubbard interaction.

\begin{figure}[tbp]
\begin{center}
\includegraphics[height=.4\textheight]{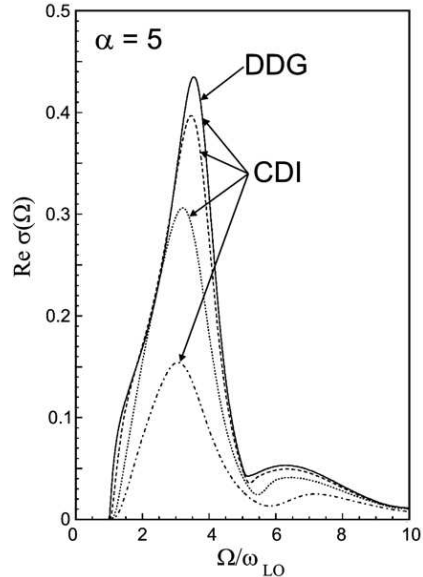}
\end{center}
\caption{Optical 
conductivity of a polaron gas at $T=0$ as a function of the frequency as calculated in Ref. \protect\cite{catau2} (CDI) for different values
of the electron density: $n_{0}=1.4\times 10^{-5}$ (solid curve), $n_{0}=1.4\times 10^{-4}$ (dashed curve), 
$n_{0}=1.4\times 10^{-3}$ (dotted curve), and $n_{0}=1.4\times 10^{-2}$ (dotted-dashed curve). 
The electron density is measured per $R_{p}^{3}$, where $R_{p}$ is the Fr\"{o}hlich polaron radius. 
The value of $\protect\varepsilon_{0}/\protect\varepsilon_{\infty }$ is 3.4.
The solid curve practically coincides with the {known} 
optical conductivity of a single polaron \protect\cite{DSG72} 
(DDG). 
(After Ref. \protect\cite{catau2}.)
}
\label{fig:P24-12}
\end{figure}

The optical conductivity of a many-polaron gas {was} further investigated, by Iadonisi and associates,
e.~g. in Ref. \cite{catau2} within the memory-function approach \cite{DSG72} by
calculating the correction to the dielectric function of the electron gas,
due to the electron-phonon interaction with variational parameters of a
single-polaron Feynman model. A suppression of the optical absorption from
the one-polaron optical absorption of Ref.\,\cite{DSG72} with
increasing density shown in Fig. \ref{fig:P24-12} is expected because of the screening of the Fr\"{o}hlich
interaction with increasing polaron density.

Recently, dynamical exchange effects were found to lead to noticeable corrections to the 
random-phase approximation results for the ground state energy, the effective mass, and 
the optical conductivity of a two-dimensional many-polaron gas in Refs. \cite{Hameeuw2003,Hameeuw2004}.

\subsection{Many-Polaron Cyclotron Resonance in Quantum Wells}

\begin{figure}[tbp]
\begin{center}
\includegraphics[height=.4\textheight]{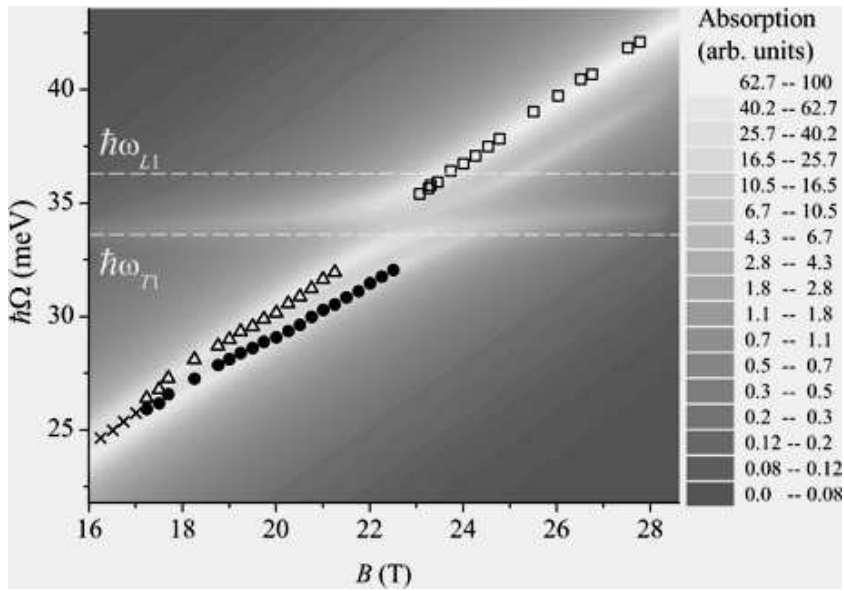}
\end{center}
\caption{Density map of the magnetoabsorption spectrum for a 10-nm GaAs/AlAs
quantum well as calculated in Ref.~\protect\cite{SPRB-2}. Peak positions of 
the experimental spectra are taken from Fig.\,3 of Ref. \protect\cite{Poulter2001}. 
Crosses indicate the CR peak positions for $B\leq17$ T. Above 17 T, the CR line 
splits into two components represented by triangles and by {filled} dots. 
{This splitting occurs due to the fact, that the electron Landau levels in 
a non-parabolic conduction band are non-equidistant. The higher- and lower-energy 
components correspond to the transitions between the Landau levels (0 $\to$ 1) and (1 $\to$ 2), 
respectively.} A single CR line shown by squares is observed above 23 T. Dashed lines show LO- and TO-phonon 
energies in GaAs. (After Ref.~\protect\cite{SPRB-2}.)}
\label{fig:P24-7}
\end{figure}

In GaAs/AlAs quantum wells with high electron density, anticrossing of the cyclotron-resonance spectra 
{has been observed} near the GaAs transverse optical (TO) phonon frequency $\omega _{T1}$ rather than near the 
GaAs LO-phonon frequency $\omega _{L1}$ \cite{Poulter2001}. {This anticrossing near $\omega _{T1}$ was explained in the framework of the polaron theory in Refs. \cite{SPRB-2,Comment} (Fig.\thinspace \ref{fig:P24-7})}. 
This effect is in contrast with the cyclotron resonance of a low-density polaron gas in a quantum well, 
where anticrossing occurs near the LO-phonon frequency. The appearance 
of the anticrossing frequency close to $\omega_{T1}$ instead of $\omega _{L1}$ is due to the screening of 
the electron-phonon interaction by the plasma {oscillations}. The theory takes into account 
the magnetoplasmon-phonon mixing and the electron-phonon interaction 
with all the phonon modes specific for the quantum well. A discussion on alternative explanations of the data 
of Ref. \cite{Poulter2001} is still going on \cite{Faugeras2004,Comment,Faugeras2005}.

\subsection{Interacting Polarons in a Quantum Dot}

A system of $N$ electrons (or holes) with mutual Coulomb repulsion and interacting with the bulk phonons 
is analysed in Refs. \cite{SSC114-305,KFBD2004}. It has been shown \cite{PRE96} that the path-integral approach 
to the many-body problem for a fixed number of identical particles can be formulated as a
Feynman-Kac functional on a state space for $N$ indistinguishable particles.
The resulting variational inequality for identical particles \cite{DevreeseFluc} 
formally has the same structure as the Jensen-Feynman variational principle \cite{Feynman}.
A model system consisting of $N$ electrons and $N_{f}$ fictitious particles 
in a harmonic confinement potential with elastic interparticle interactions 
is studied in Refs. \cite{SSC114-305,KFBD2004}. The Lagrangian of this model system is symmetric 
with respect to electron permutations. The parameters of the model system are found from the variational 
principle \cite{DevreeseFluc} and used for the calculation of the ground-state energy and the optical 
conductivity of the $N$-polaron system in a quantum dot \cite{SSC114-305,KFBD2004}.

The polaron contribution to the ground-state energy of the $N$-polaron system in a quantum dot per 
particle calculated within our variational path-integral treatment for different $N$ 
\cite{BKD2005} is compared with the respective quantity for a polaron gas in bulk, 
found in Ref. \cite{FCI1999} within an intermediate-coupling approach 
and in Ref. \cite{CS1993}, using a variational approach developed first in Ref. \cite{LDB77}. Our all-coupling
variational method \cite{BKD2005} provides somewhat lower values for the polaron contribution than
those obtained in Refs. \cite{FCI1999,CS1993} and the trends predicted in those references are confirmed.

\clearpage\begin{figure}[tbh]
\begin{center}
\includegraphics[width=0.8\textwidth]{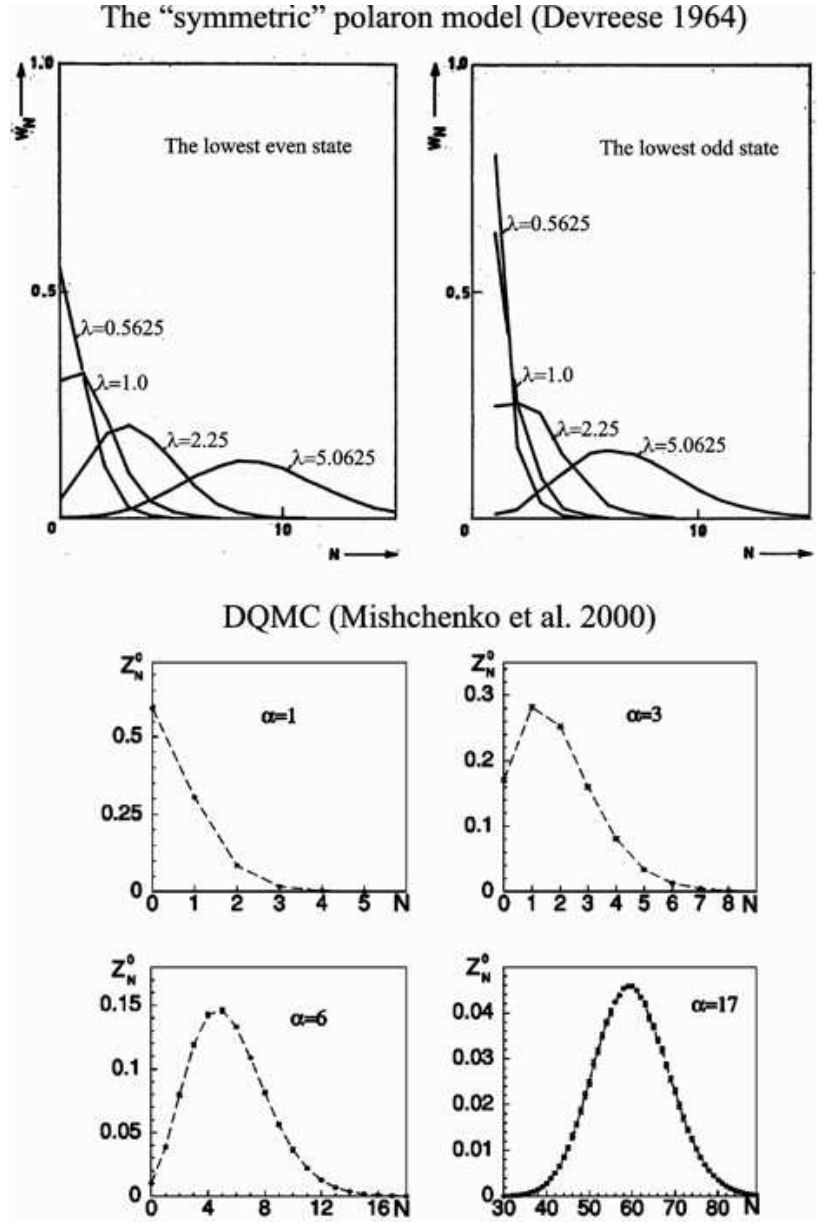} \end{center}
\caption{\textit{Upper panel}: The phonon distribution functions $W_{N}$ in
the \textquotedblleft  symmetric\textquotedblright\ polaron model for various
values of the effective coupling constant $\lambda$ at $\varkappa=1,\mathbf{P}=0$
(from \cite{DThesis}, Fig. 23). \textit{Lower panel}: Distribution of
multiphonon states in the polaron cloud within DQMC method for various values
of $\alpha$ (from \cite{Mishchenko2000}, Fig. 7).}%
\label{NPolarons}%
\end{figure} \clearpage

\appendix 
\section {On the Contributions of the $N$-phonon States to the Polaron Ground State}
\label{section-A1}
In Ref. \cite{DThesis}, the present author introduced and analysed an exactly solvable 
(\textquotedblleft symmetric\textquotedblright) 1D-polaron model.
The further study of this model was performed in Refs. \cite{DE64,DE68}. This model can be used as a tool for the 
evaluation of the accuracy of the standard polaron theories \cite{DE63,DE65,DE66}.

The model consists of an electron interacting with two oscillators possessing 
opposite wave vectors:\textbf{\ }$\mathbf{k}$ and $\mathbf{-k}$. The parity operator, which changes 
$a_{\mathbf{k}}$ and $a_{-\mathbf{k}}$ (and also $a_{\mathbf{k}}^{\dag}$ and $a_{-\mathbf{k}}^{\dag}$), 
commutes with the Hamiltonian of the system. Hence, the polaron states are classified into even and odd 
states with eigenvalues of the parity operator $+$1 and $-$1, respectively. For the lowest even and odd states, 
the phonon distribution functions $W_{N}$ are plotted in Fig. \ref{NPolarons}, upper panel, for some values 
of the effective coupling constant $\lambda$ of this \textquotedblleft symmetric\textquotedblright model. 
The value of the parameter $\varkappa=\sqrt{\left(  \hbar k\right)^{2}/ m_{b}\hbar\omega_{\mathrm{LO}}}$ 
for these graphs was taken to be 1, while the total polaron momentum $\mathbf{P}=0$. In the weak-coupling 
case ($\lambda\approx0.6$) $W_{N}$ is a decaying function of $N$. When increasing $\lambda$, $W_{N}$ acquires 
a maximum, e.~g. at $N=8$ for the lowest even state at $\lambda=5.0625$. The phonon distribution function 
$W_{N}$ has the same character for the lowest even and the lowest odd states at all values of the number 
of the virtual phonons in the ground state (as distinct from the higher states). This led to the conclusion 
that the lowest odd state is an internal excited state of the polaron. 

In Ref. \cite{Mishchenko2000}, the structure of the Fr\"{o}hlich polaron cloud was investigated using the DQMC method. Contributions of $N$-phonon states to the polaron ground state 
were calculated as a function of $N$ for a few values of the coupling constant $\alpha,$ see Fig. \ref{NPolarons}, 
lower panel. The evolution from the weak-coupling case ($\alpha=1$) into the strong-coupling 
regime ($\alpha=17$) was studied. Comparison of the lower panel to the upper panel in Fig. \ref{NPolarons} shows that 
the evolution of the shape and the scale of the distribution of the $N$-phonon states with increasing $\alpha$ 
as derived for a Fr\"{o}hlich polaron within DQMC method \cite{Mishchenko2000} is 
\textit{in notable agreement} with the results obtained within the \textquotedblleft symmetric\textquotedblright 1D polaron model 
\cite{DThesis,DE64,DE68}.

\acknowledgments I thank V. M. Fomin for discussions of the present paper and S. N. Klimin for interactions on some 
aspects of this work. The present work has been supported by the GOA BOF UA 2000, IUAP, FWO-V projects G.0306.00, G.0274.01N, G.0435.03, 
the WOG WO.035.04N (Belgium), the European Commission GROWTH Programme, NANOMAT project, contract 
No. G5RD-CT-2001-00545 and the European Commission SANDIE Network of Excellence, contract No. NMP4-CT-2004-500101.

\end{document}